\documentclass[10pt]{article}
\usepackage{graphicx,bm}

\def\norm#1{\left|\mkern-2mu\left|#1\right|\mkern-2mu\right|}
\def\k{{\bm k}}
\def\r{{\bm r}}
\def\u{{\bm u}}
\begin{document}
\begin{center}{{\large \bf An upper bound for passive scalar diffusion 
in shear flows}\\~\\
Chuong V. Tran\\
School of Mathematics and Statistics, University of St Andrews\\
St Andrews KY16 9SS, United Kingdom}
\end{center}
This study is concerned with the diffusion of a passive scalar 
$\Theta(\r,t)$ advected by general $n$-dimensional shear flows 
$\u=u(y,z,\cdots,t)\widehat{x}$ having finite mean-square velocity 
gradients. The unidirectionality of the incompressible flows conserves 
the stream-wise scalar gradient, $\partial_x\Theta$, allowing only the 
cross-stream components to be amplified by shearing effects. This 
amplification is relatively weak because an important contributing 
factor, $\partial_x\Theta$, is conserved, effectively rendering a slow 
diffusion process. It is found that the decay of the scalar variance 
$\langle\Theta^2\rangle$ satisfies 
$d\langle\Theta^2\rangle/dt\ge -C\kappa^{1/3}$, where $C>0$ is a 
constant, depending on the fluid velocity gradients and initial 
distribution of $\Theta$, and $\kappa$ is the molecular diffusivity. 
This result generalizes to axisymmetric flows on the plane and on 
the sphere having finite mean-square angular velocity gradients.

~

\centerline{* * * * *}

The transport of a diffusive scalar $\Theta(\r,t)$ by incompressible 
fluid flows $\u(\r,t)$ is governed by
\begin{eqnarray}
\label{gov}
(\partial_t + \u\cdot\nabla - \kappa\Delta)\Theta &=& 0, \\
\nabla\cdot\u &=& 0, \nonumber 
\end{eqnarray}
where $\kappa$ is the molecular diffusivity. This system has been an 
active research subject$^{1-10}$ for its application to problems in 
geophysical, environmental, and industrial context. Various types of 
flows have been considered, ranging from single-scale$^{6,9}$ to fully 
turbulent flows.$^{2,7}$ A problem of genuine interest has been the 
decay behaviour of a finite initial scalar distribution in the limit 
of small $\kappa$, where the question is whether the decay rate 
$d\langle\Theta^2\rangle/dt$ of the scalar variance 
$\langle\Theta^2\rangle$ remains nonzero as $\kappa\to0$. In two 
dimensions, recent results$^{1,7}$ have suggested a negative answer to 
this question. However, one of these results$^1$ was derived for finite 
time only, while the other result$^7$ relied on an assumption of 
power-law spectra of $\langle\Theta^2\rangle$.

For shear flows, the advection-diffusion problem is more tractable, even 
in higher dimensions, making it possible for the above answer to be 
derived rigorously, as is presently shown. Moreover, in this case, the 
decay rate approaches zero relatively rapidly as $\kappa\to0$. The main 
feature in shear flows, not shared by turbulent flows, that facilitates 
this derivation is that shear flows conserve the stream-wise scalar 
gradient, say $\partial_x\Theta$, allowing only the cross-stream 
components to be amplified. The amplification of these components is 
non-accelerated since an important contributing factor, $\partial_x\Theta$, 
is conserved. More quantitatively, shearing effects can give rise to a 
quadratic growth of the mean-square scalar gradients 
$\langle|\nabla\Theta|^2\rangle$, as opposed to exponential growth, 
presumably occurring for turbulent flows. This slow growth of 
$\langle|\nabla\Theta|^2\rangle$ is overwhelmingly suppressed by 
diffusive effects, even in the limit $\kappa\to0$. As a consequence, 
given a smooth initial scalar distribution, the maximally achievable 
value of $\langle|\nabla\Theta|^2\rangle$ grows relatively slowly as 
$\kappa$ is decreased. This results in a rapid approach of the decay
rate $\kappa\langle|\nabla\Theta|^2\rangle$ to zero as $\kappa\to0$.

This Brief Communication derives an upper bound for the decay rate 
$\kappa\langle|\nabla\Theta|^2\rangle$ for general $n$-dimensional 
shear flows $\u=u(y,z,\cdots,t)\widehat{x}$. For simplicity, periodic 
boundary conditions are considered, and the flows are assumed to have 
finite velocity gradients or just finite mean-square velocity gradients. 
The initial scalar distribution is assumed to have finite mean-square
gradients. The derived upper bound for $\kappa\langle|\nabla\Theta|^2\rangle$ 
is found to scale as $\kappa^{1/3}$. This analytic result is valid 
uniformly in time and generalizes to axisymmetric flows on the plane and 
on the sphere having integrable differential angular velocity.

For $\u=u(y,z,\cdots,t)\widehat{x}$, equation (\ref{gov}) becomes
\begin{eqnarray}
\label{gov-shear}
(\partial_t + u\,\partial_x - \kappa\Delta)\Theta = 0.
\end{eqnarray}
As $u(y,z,\cdots,t)$ is independent of the flow direction $x$, the 
classes of Fourier modes $\widehat\Theta(\k,t)$ 
having a common $k_x$ are dynamically decoupled.$^9$ This is directly 
connected to the fact that the derivatives $\partial^m_x\Theta$, for 
$m=1,2,3,\cdots$, are materially conserved, as can be seen from the
governing equation for $\partial^m_x\Theta$,
\begin{eqnarray}
(\partial_t + u\,\partial_x - \kappa\Delta)\partial^m_x\Theta = 0.
\end{eqnarray}
This advection-diffusion equation is the same as that for $\Theta$;
hence, all conservation laws for $\Theta$ also apply to 
$\partial^m_x\Theta$. In particular, the supremum
$\norm{\partial_x\Theta}_\infty$ and the mean-square stream-wise
scalar gradient $\langle|\partial_x\Theta|^2\rangle$, both being used 
in the subsequent calculations, are conserved by the advection term. 
Under diffusive effects, these quantities decay in time, and hence 
are bounded by their initial values.

The approach of Tran$^{11}$ (see also Tran and Dritschel$^{7}$) for
estimating the enstrophy dissipation in two-dimensional turbulence is 
now applied to the present case. As the evolution of $\partial_x\Theta$ 
is trivial, one can drop this component from consideration. However, 
there appears to be no gain for so doing. Hence, for clarity, all the 
subsequent calculations make no separation of $\partial_x\Theta$ from
$\nabla\Theta$. The evolution equation for $\nabla\Theta$ is
\begin{eqnarray}
\label{gov-sheary}
(\partial_t + u\,\partial_x - \kappa\Delta)\nabla\Theta = 
-\partial_x\Theta\,\nabla u.
\end{eqnarray}
The `forcing' term on the right-hand side of (\ref{gov-sheary}) depends
on the velocity gradient $\nabla u$ and on the decaying stream-wise 
gradient $\partial_x\Theta$, but does not depend on the cross-stream 
gradients. This means that each cross-stream component is amplified 
independently and that $\langle|\nabla\Theta|^2\rangle$ cannot grow more 
rapidly than quadratic in time. It is also worth mentioning that since 
$\partial_x\Theta$ can only decay, an initial scalar distribution 
homogeneous in the flow direction, i.e. $\partial_x\Theta=0$, decays purely 
diffusively because the right-hand side of (\ref{gov-sheary}) is identically
zero for all $t>0$. In other words, shearing effects alone cannot amplify 
any cross-stream scalar gradients in the absence of $\partial_x\Theta$. 
This is obvious from physical point of view. The evolution equation for 
$\langle|\nabla\Theta|^2\rangle$ is obtained by multiplying (\ref{gov-sheary}) 
by $\nabla\Theta$ and taking the spatial average of the resulting equation,
\begin{eqnarray}
\label{diff}
\frac{1}{2}\frac{d}{dt}\langle|\nabla\Theta|^2\rangle 
&=&
-\langle\partial_x\Theta\,\nabla u\cdot\nabla\Theta\rangle 
- \kappa\langle|\Delta\Theta|^2\rangle,
\end{eqnarray}
where the advection term identically vanishes. 

Two estimates of the triple-product term in (\ref{diff}) are
obtained by using the Cauchy-Schwarz inequality:
\begin{eqnarray}
\label{estimate}
\frac{1}{2}\frac{d}{dt}\langle|\nabla\Theta|^2\rangle 
&\le& 
\cases{\langle|\nabla u|^2\rangle^{1/2}\norm{\partial_x\Theta}_\infty
\langle|\nabla\Theta|^2\rangle^{1/2} - \kappa\langle|\Delta\Theta|^2\rangle,\cr
\norm{\nabla u}_\infty\langle|\partial_x\Theta|^2\rangle^{1/2}
\langle|\nabla\Theta|^2\rangle^{1/2} - \kappa\langle|\Delta\Theta|^2\rangle.\cr}\end{eqnarray}
Either equation of (\ref{estimate}) can be used to deduce an upper 
bound for $\kappa\langle|\nabla\Theta|^2\rangle$, depending on which 
estimate of the triple-product term is more optimal. For the first 
equation, applying the Cauchy--Schwarz inequality 
$\langle|\Delta\Theta|^2\rangle\ge\langle|\nabla\Theta|^2
\rangle^2/\langle\Theta^2\rangle$ yields
\begin{eqnarray}
\label{estimate1}
\frac{1}{2}\frac{d}{dt}\langle|\nabla\Theta|^2\rangle 
&\le& \langle|\nabla u|^2\rangle^{1/2}\norm{\partial_x\Theta}_\infty
\langle|\nabla\Theta|^2\rangle^{1/2}
- \kappa\frac{\langle|\nabla\Theta|^2\rangle^2}{\langle\Theta^2\rangle}\\
&\le&
\frac{\langle|\nabla\Theta|^2\rangle^{1/2}}{\langle\Theta^2\rangle}
\left(\langle|\nabla u|^2\rangle^{1/2}\norm{\partial_x\Theta}_\infty
\langle\Theta^2\rangle - \kappa\langle|\nabla\Theta|^2\rangle^{3/2}\right).
\nonumber
\end{eqnarray}
The first term in the brackets on the right-hand side of (\ref{estimate1})
depends on $\langle|\nabla u|^2\rangle$ (which is equivalent to 
$\langle|\nabla\times\u|^2\rangle$ in this case) and on the decaying 
quantities $\norm{\partial_x\Theta}_\infty$ and $\langle\Theta^2\rangle$. 
Given a bounded $\langle|\nabla u|^2\rangle$ and bounded initial 
$\norm{\partial_x\Theta}_\infty$ and $\langle\Theta^2\rangle$, 
this term remains bounded. It follows that for $t>0$,
\begin{eqnarray}
\label{estimate2}
\kappa\langle|\nabla\Theta|^2\rangle^{3/2} &\le& c,
\end{eqnarray}
provided that it holds for $t=0$. Here, $c$ is an upper bound 
for $\langle|\nabla u|^2\rangle^{1/2}\norm{\partial_x\Theta}_\infty
\langle\Theta^2\rangle$. From (\ref{estimate2}) one can readily deduce that 
\begin{eqnarray}
\label{estimate3}
\kappa\langle|\nabla\Theta|^2\rangle &\le& c^{2/3}\kappa^{1/3} = 
\frac{C}{2}\kappa^{1/3},
\end{eqnarray}
where $C$ is an upper bound for $2\langle|\nabla u|^2\rangle^{1/3}\norm{
\partial_x\Theta}_\infty^{2/3}\langle\Theta^2\rangle^{2/3}$. 
The decay of $\langle\Theta^2\rangle$ then satisfies
\begin{eqnarray}
\label{estimate4}
\frac{d}{dt}\langle\Theta^2\rangle 
&=& -2\kappa\langle|\nabla\Theta|^2\rangle \ge - C\kappa^{1/3}.
\end{eqnarray}

The same bound but with a different constant $C$ can be obtained by 
manipulating the second equation of (\ref{estimate}) along these lines. 
In this case, $C$ denotes an upper bound for $2\norm{\nabla u}_\infty^{2/3}
\langle|\partial_x\Theta|^2\rangle^{1/3}\langle\Theta^2\rangle^{2/3}$,
and instead of the requirement $\langle|\nabla u|^2\rangle<\infty$, the 
slightly more stringent condition $\norm{\nabla u}_\infty<\infty$ is assumed.

Equation (\ref{estimate4}) implies a slow decay of 
$\langle\Theta^2\rangle$ in the limit of small $\kappa$. In other 
words, shear flows are rather ineffective mixers. The slow decay of 
$\langle\Theta^2\rangle$ has a bearing on its exponential decay rate, 
which is of particular interest and has been widely studied. For handling 
exponential decay behaviour, equation (\ref{estimate4}) can be 
rewritten in the more convenient form,
\begin{eqnarray}
\label{estimate5}
\frac{1}{\langle\Theta^2\rangle}\frac{d}{dt}\langle\Theta^2\rangle 
&\ge& -\frac{C\kappa^{1/3}}{\langle\Theta^2\rangle}.
\end{eqnarray}
Suppose that at any instance in time, the decay of 
$\langle\Theta^2\rangle$ is approximated by a pure exponential decay 
at the rate of $\lambda$, then by (\ref{estimate5}), $\lambda$ satisfies
\begin{eqnarray}
\label{estimate6}
\lambda &\le& \frac{C\kappa^{1/3}}{\langle\Theta^2\rangle}.
\end{eqnarray}
This equation gives an explicit upper bound for $\lambda$ in terms of
$\kappa$. 

During the period (or periods) of scalar gradient growth, i.e.
$d\langle|\nabla\Theta|^2\rangle/dt\ge0$, the constant $C$ in 
(\ref{estimate6}) need not be an upper bound for the quantity in
question, but can be its instantaneous value (cf. (\ref{estimate})
and the subsequent calculations). Therefore, in such a period, $\lambda$
satisfies
\begin{eqnarray}
\label{estimate7}
\lambda=2\kappa\frac{\langle|\nabla\Theta|^2\rangle}{\langle\Theta^2\rangle}
&\le& 
\cases{2\langle|\nabla u|^2\rangle^{1/3}\norm{\partial_x\Theta}_\infty^{2/3}
\langle\Theta^2\rangle^{-1/3}\kappa^{1/3} ,\cr
2\norm{\nabla u}_\infty^{2/3}\langle|\partial_x\Theta|^2\rangle^{1/3}
\langle\Theta^2\rangle^{-1/3}\kappa^{1/3}.\cr}
\end{eqnarray}
For the period (or periods) of scalar gradient decay, i.e.
$d\langle|\nabla\Theta|^2\rangle/dt<0$, the validity of (\ref{estimate7}) 
could become questionable only if $\langle|\nabla\Theta|^2\rangle$ has 
decayed more slowly than $\langle\Theta^2\rangle$, i.e. the small scales 
have decayed relatively more slowly than the large scales. This condition
requires that the production of cross-stream scalar gradients remain 
considerable throughout their decay. In any case, from (\ref{estimate6}) 
one can conclude that in the limit of small $\kappa$, $\lambda$ could 
become sizable only when $\langle\Theta^2\rangle\propto\kappa^{1/3}$, 
i.e. when the scalar distribution has become virtually homogeneous.
Furthermore, the remaining fraction of $\langle\Theta^2\rangle$ would then
have been `cascading' to a dissipation wavenumber, say $k_d$, satisfying 
$k_d\propto\kappa^{-1/2}$. It is notable that for a simple shear flow in 
two dimensions satisfying all the required conditions, Vanneste and 
Byatt--Smith$^{12}$ argue that fast decay of the scalar is possible, in 
the sense that $\lambda$ remains nonzero in the limit $\kappa\to0$. The 
present result implies that the asymptotic regime for considering this 
possibility is $\langle\Theta^2\rangle\propto\kappa^{1/3}$.

The present result readily generalizes to axisymmetric flows on the plane 
(vortical flows) and on the sphere (zonal flows), provided that these flows 
do not become singular, in the sense to be described in due course. For 
these cases, the respective advection-diffusion equations in the plane
polar and spherical polar coordinates~are
\begin{eqnarray}
\label{polar}
\left(\partial_t + \Omega(r,t)\,\partial_\phi - \kappa(r^{-1}\,\partial_r
(r\partial_r) + r^{-2}\,\partial^2_\phi)\right)\Theta = 0
\end{eqnarray}
and
\begin{eqnarray}
\label{spherical}
\left(\partial_t + \Omega(\theta,t)\,\partial_\phi - \kappa
(\sin^{-1}\theta\,\partial_\theta(\sin\theta\,\partial_\theta)
 + \sin^{-2}\theta\,\partial^2_\phi)\right)\Theta = 0.
\end{eqnarray}
Here $\Omega(\cdot,t)$ denotes the fluid angular velocity, and all other 
notations are standard. The sphere radius has been set to unity. For 
(\ref{polar}) zero boundary conditions are imposed. Similar to the previous 
case, the stream-wise scalar gradient in each case ($\partial_\phi\Theta/r$ 
for vortical flows and $\partial_\phi\Theta/\sin\theta$ for zonal flows) 
is conserved by the flows. However, these are not known to decay under 
diffusive effects. Instead, the decaying quantities in these cases are the 
derivatives $\partial^m_\phi\Theta$, which are governed by the same equation 
as $\Theta$ in each case. Now it is straightforward to perform the above 
calculations in curvilinear coordinates. For vortical flows, one obtains
\begin{eqnarray}
\label{diff1}
\frac{1}{2}\frac{d}{dt}\langle|\nabla\Theta|^2\rangle 
&=& \langle\Delta\Theta\,\Omega\,\partial_\phi\Theta\rangle 
- \kappa\langle|\Delta\Theta|^2\rangle 
=
-\langle\partial_r\Theta\,\partial_r\Omega\,\partial_\phi\Theta\rangle
- \kappa\langle|\Delta\Theta|^2\rangle \nonumber\\
&\le&
\langle|\partial_r\Theta|^2\rangle^{1/2}
\langle|\partial_r\Omega|^2\rangle^{1/2}
\norm{\partial_\phi\Theta}_\infty
- \kappa\langle|\Delta\Theta|^2\rangle\\
&\le&
\frac{\langle|\nabla\Theta|^2\rangle^{1/2}}{\langle\Theta^2\rangle}
\left(\langle|\partial_r\Omega|^2\rangle^{1/2}\norm{\partial_\phi\Theta}_\infty
\langle\Theta^2\rangle - \kappa\langle|\nabla\Theta|^2\rangle^{3/2}\right).
\nonumber
\end{eqnarray}
Similarly for zonal flows, one obtains
\begin{eqnarray}
\label{diff2}
\frac{1}{2}\frac{d}{dt}\langle|\nabla\Theta|^2\rangle 
&=& \langle\Delta\Theta\,\Omega\,\partial_\phi\Theta\rangle 
- \kappa\langle|\Delta\Theta|^2\rangle 
=
-\langle\partial_\theta\Theta\,\partial_\theta\Omega\,
\partial_\phi\Theta\rangle
- \kappa\langle|\Delta\Theta|^2\rangle \nonumber\\
&\le&
\langle|\partial_\theta\Theta|^2\rangle^{1/2} 
\langle|\partial_\theta\Omega|^2\rangle^{1/2}
\norm{\partial_\phi\Theta}_\infty
- \kappa\langle|\Delta\Theta|^2\rangle\\
&\le&
\frac{\langle|\nabla\Theta|^2\rangle^{1/2}}{\langle\Theta^2\rangle}
\left(\langle|\partial_\theta\Omega|^2\rangle^{1/2}
\norm{\partial_\phi\Theta}_\infty\langle\Theta^2\rangle - 
\kappa\langle|\nabla\Theta|^2\rangle^{3/2}\right).
\nonumber
\end{eqnarray}
From (\ref{diff1}) and (\ref{diff2}), one can recover the result 
$d\langle\Theta^2\rangle/dt \ge -C\kappa^{1/3}$ derived earlier for the 
periodic case. The constant $C$ is an upper bound for 
$2\langle|\partial_r\Omega|^2\rangle^{1/3}
\norm{\partial_\phi\Theta}_\infty^{2/3}\langle\Theta^2\rangle^{2/3}$ 
and
$2\langle|\partial_\theta\Omega|^2\rangle^{1/3}
\norm{\partial_\phi\Theta}_\infty^{2/3}\langle\Theta^2\rangle^{2/3}$ for 
the vortical and zonal cases, respectively. Here 
$\norm{\partial_\phi\Theta}_\infty$ 
plays the role of $\norm{\partial_x\Theta}_\infty$ and
$\langle|\partial_r\Omega|^2\rangle$ and 
$\langle|\partial_\theta\Omega|^2\rangle$ 
play the role of 
$\langle|\nabla u|^2\rangle$. The condition for the flows is
the integrability of the differential rotation 
$\langle|\partial_r\Omega|^2\rangle<\infty$ and
$\langle|\partial_\theta\Omega|^2\rangle<\infty$, rather than the 
integrability of the velocity gradients. Note that for the case of 
vortical flows, Bajer, Bassom, and Gilbert$^{13}$ have found by a 
different method that the decay rate 
$\kappa\langle|\nabla\Theta|^2\rangle$ scales as $k^{1/3}$ 
(also see Rhines and Young$^{14}$).

In conclusion, this Brief Communication has derived a rigorous upper 
bound for the decay rate of the variance $\langle\Theta^2\rangle$ 
of a passive scalar $\Theta$ in general $n$-dimensional shear flows. 
The flows are assumed to have finite velocity gradients or just finite 
mean-square velocity gradients, and the initial scalar distribution is 
assumed to be smooth. This upper bound is valid uniformly in time and 
scales as $\kappa^{1/3}$, where $\kappa$ is the diffusivity. This implies 
that in the limit of small diffusivity, the diffusion of a passive scalar 
in shear flows is slow: shear flows are rather poor mixers. The reason is 
that shear flows conserve the stream-wise scalar gradient and amplify the 
cross-stream scalar gradients relatively weakly. This result generalizes 
to axisymmetric flows on the plane and on the sphere having finite 
mean-square angular velocity gradients.

The author would like to thank Prof David Dritschel for bringing to his 
attention the paper of Bajer, Bassom, and Gilbert$^{13}$ and the extension 
from the Cartesian to curvilinear cases. He would also like to acknowledge
helpful discussions with Prof Christos Vassilicos concerning the pathological
case of vortical flows, where the fluid angular velocity becomes singular
at the origin.$^{15}$ He is grateful to Prof Raymond Pierrehumbert and 
an anonymous referee for comments, which were helpful in improving this 
manuscript. 

~

\centerline{* * * * *}

~

\noindent 1. R. J. Di Perna and P. L. Lions, ``Ordinary differential
equations, transport theory and Sobolev spaces,'' Invent. Maths.
{\bf 98}, 511 (1989).

\noindent 2. D. R. Fereday and P. H. Haynes, ``Scalar decay in 
two-dimensional chaotic advection and Batchelor-regime turbulence,'' 
Phys. Fluids {\bf 16}, 4359 (2004). 

\noindent 3. P. H. Haynes and J. Vanneste, ``What controls the decay rate
of passive scalar in smooth random flows?'' Phys. Fluids {\bf 17}, 097103 
(2005). 

\noindent 4. W. Liu, ``Does a fast mixer really exist?'' Phys. Rev. E
{\bf 72}, 016312 (2005). 

\noindent 5. R. T. Pierrehumbert, ``Tracer microstructure in the large-eddy
dominated regime,'' Chaos, Solitons Fractals. {\bf 4}, 1091 (1994).

\noindent 6. J. Sukhatme and R. T. Pierrehumbert, ``Decay of passive scalars
under the action of single scale smooth velocity fields in bounded 
two-dimensional domains: from non-self-similar probability distribution 
function to self-similar Eigenmodes,'' Phys. Rev. E {\bf 66}, 056302 (2002).

\noindent 7. C. V. Tran and D. G. Dritschel, ``Vanishing enstrophy
dissipation in two-dimensional Navier--Stokes turbulence in the inviscid
limit,'' J. Fluid Mech. {\bf 559}, 107 (2006).

\noindent 8. Y. K. Tsang, T. M. Antonson, and E. Ott, ``Exponential decay
of chaotically advected passive scalars in the zero diffusivity limit,''
Phys. Rev. E {\bf 71} 066301 (2005).

\noindent 9. J. Vanneste, ``Intermittency of passive-scalar decay: Strange
eigenmodes in random shear flows,'' Phys. Fluids {\bf 18}, 087108 (2006).

\noindent 10. A. D. Majda, ``The random uniform shear layer: an explicit
example of turbulent diffusion with broad tail probability distributions,'' 
Phys. Fluids A {\bf 5}, 1963 (1993). 

\noindent 11. C. V. Tran, ``Enstrophy dissipation in freely evolving
two-dimensional turbulence,'' Phys. Fluids {\bf 17}, 081704 (2005). 

\noindent 12. J. Vanneste and J. G. Byatt--Smith, ``fast scalar decay in a 
shear flow: modes and pseudomodes'' J. Fluid Mech. {\bf 572}, 219 (2007).

\noindent 13. K. Bajer, A. P. Bassom, and A. D. Gilbert, ``Accelerated
diffusion in the centre of a vortex,'' J. Fluid Mech. {\bf 437}, 395 (2001).

\noindent 14. P. B. Rhines and W. R. Young, ``How rapidly is a passive
scalar mixed within closed streamlines?'' J. Fluid Mech. {\bf 133}, 133 (1982).

\noindent 15. P. Flohr and J. C. Vassilicos, ``Accelerated scalar dissipation
in a vortex'' J. Fluid Mech. {\bf 348}, 295 (1997).

\end{document}